\documentclass[aps,prc,twocolumn,noshowpacs, superscriptaddress,floatfix]{revtex4}

\usepackage{amsmath}
\usepackage{amssymb}
\usepackage{graphicx}
\usepackage{dcolumn}
\usepackage{bm}
\usepackage{color}
\usepackage{ulem}

\begin{document}


\title{New algorithm in the variation after projection calculations for non-yrast nuclear states }

\author{ Jia-Qi Wang }
\affiliation{China Institute of Atomic Energy, Beijing 102413,
People's Republic of China}
\affiliation{College of Physics, Jilin University, Changchun 130012, China}

\author{ Zao-Chun Gao}
\email{zcgao@ciae.ac.cn}
\affiliation{China Institute of Atomic Energy, Beijing 102413,
People's Republic of China}

\author{ Ying-Jun Ma }
\affiliation{College of Physics, Jilin University, Changchun 130012, China}

\author{Y. S. Chen}
\affiliation{China Institute of Atomic Energy, Beijing 102413,
People's Republic of China}
\date{\today}

\begin{abstract}
We present a novel and simple algorithm in the variation after projection (VAP) approach for the non-yrast nuclear states. It is for the first time that the yrast state and non-yrast states can be varied on the same footing. The orthogonality among the calculated states is automatically fulfilled by solving the Hill-Wheeler equation. This avoids the complexity of the frequently used Gram-Schmidt orthogonalization, as adopted by the excited VAMPIR method. Thanks to Cauchy's interlacing theorem in matrix theory, the sum of the calculated lowest projected energies with the same quantum numbers can be safely minimized. Once such minimization is converged, all the calculated energies and corresponding states can be obtained, simultaneously. The present VAP calculations are performed with time-odd Hartree-Fock Slater determinants. It is shown that the calculated VAP energies (both yrast and non-yrast) are very close to the corresponding ones from the full shell model calculations. It appears the present algorithm is not limited to the VAP, but should be universal, i.e., one can do the variation with different forms of the many-body wavefunctions to calculate the excited states in different quantum many-body systems.
\end{abstract}

\maketitle

For a given many-body quantum system, its spectrum and the corresponding wave functions should be obtained by solving Schr\"{o}dinger's equation. In nuclear structure studies, this is done in the full shell model(SM) calculations with a given Hamiltonian. However, due to the combinatorial computational cost, full SM calculations up to now have been restricted to rather small model spaces. To tackle the problem of the eigensystems in even larger model spaces, it looks that the only way is to compress the huge configuration space into a relatively small one so that the code can be run on a present-day computer. Unfortunately, the obtained energies and states are approximated ones. To make such approximated solutions as good as possible, various methods, such as shell model truncation\cite{Horoi94}, stochastic quantum Monte Carlo approaches \cite{Koonon97,Otsuka01}, and the VAMPIR method \cite{Schmid04} have been developed.

Recently, we implemented a variation after projection (VAP) calculation \cite{tuya17}. Instead of using the Hartree-Fock-Bogoliubov (HFB) vacuum state as adopted in the VAMPIR approach, we take a time-odd Hartree-Fock(HF) Slater determinant(SD). This means the particle number projection can be omitted, and only the spin projection is required. Although the VAP wave function is simply a spin projected SD, our calculations show that it can still be very close to the corresponding shell model wave function. This is because the involved spin projection plays a key role in obtaining a good shell model approximation\cite{gao15,tuya17}. However, that VAP can only be applied to the yrast states. In the present work, we construct a new and easy algorithm to extend our VAP calculations to the non-yrast excited states and show that the calculated non-yrast states are also very close to the exact shell model ones.

We should mention that the EXCITED VAMPIR can be used to calculate the excited states, too. In that method, one should first vary the HFB vacuum state to find the energy minimum for the yrast state. Then one varies the second HFB vacuum for the first excited state, but the Gram-Schmidt orthogonalization must be applied at each iteration to ensure the orthogonality between the first excited state and the yrast one. Similarly, the HFB vacua are added and varied one by one for the higher excited states, and the Gram-Schmidt orthogonalization should be treated through out all the VAMPIR iterations. To incorporate the important correlations, the more general EXCITED FED VAMPIRE (EFV) \cite{Schmid89,Schmid04} uses several instead of only one symmetry-projected HFB vacuum for the description of each state. Those symmetry-projected HFB vacua for the same nuclear state are still added and varied successively. All the VAMPIR calculations then are ended with a final diagonalization of the given Hamiltonian in the space spanned by all the obtained projected HFB vacua.

The present VAP looks similar to the EFV in the sense that both of them take the projected basis. But here, we developed a quite different strategy in optimizing the low-lying nuclear states. The projected SDs are varied simultaneously and all the calculated states with the same spin (and parity if necessary) can be obtained on the same footing. The orthogonality among the calculated states is automatically fulfilled by solving the Hill-Wheeler equation. Thus, we do not need the Gram-Schmidt orthogonalization here.

To address the present algorithm, let us first introduce Cauchy's interlacing theorem relating to the eigenvalues of Hermitian matrices (Theorem 4.3.17 in \cite{mt}). Let $A_n$ and $B_m$ be Hermitian matrices of orders $n$ and $m$, respectively. The eigenvalues of $A_n$ are denoted by $\mu_1\leq\mu_2\leq\cdots\leq\mu_{n}$. Those of $B_m$ are $\lambda_1\leq\lambda_2\leq\cdots\leq\lambda_m$.
 if $B_{n-1}$ is of order $m=n-1$ and $A_n$ is
    \begin{equation}\label{herm}
A_n
=\left(\begin{array}{cc}
B_{n-1}&\vec{y}\\\vec{y}^\dagger&a
\end{array}\right),
\end{equation}
where, $a$ is a real number and $\vec{y}$ is a complex vector of order $n-1$.
Then Cauchy's interlacing theorem tells us that
\begin{eqnarray}
\mu_1\leq\lambda_1\leq\cdots\leq\mu_j\leq\lambda_j\leq\mu_{j+1}\leq\cdots\leq\lambda_{n-1}\leq\mu_n.
\end{eqnarray}
One can immediately have the following conclusion  by repeated application of the above inequalities. For any principle submatrix $B_m$ of $A_n$, we have (Theorem 4.3.28.in \cite{mt})
\begin{eqnarray}\label{ineq2}
\mu_j\leq\lambda_j\leq\mu_{j+n-m}  \quad(1\leq j \leq m).
\end{eqnarray}

Since a unitary transformation does not change the eigenvalues of a given Hermitian matrix, the above conclusion can be naturally generalized to the following Poincar\'{e} separation theorem. Supposing that $\vec{u}_1,\vec{u}_2,...\vec{u}_m$ are orthonormal complex vectors in the $n$-dimensional space where $A_n$ is defined, the matrix elements $(B_m)_{ij}=\vec{u}^\dagger_i(A_n)\vec{u}_j$ form a Hermitian matrix $B_m$, and one can have the same inequalities (\ref{ineq2}) from Poincar\'{e} separation theorem(see Theorem 4.3.37 in \cite{mt})

However, if $\vec{u}_1,\vec{u}_2,...\vec{u}_m$ are not orthonormal but independent (If not, we remove the redundant ones), for instance, the projected states used here, one may solve the following generalized eigenvalue equation of order $m$,
\begin{eqnarray}\label{ge}
\sum_{j=1}^m[(B_m)_{ij}-\lambda_k (N_m)_{ij}]f^k_j=0,
\end{eqnarray}
where $(B_m)_{ij}=\vec{u}^\dagger_i(A_n)\vec{u}_j$ and $(N_m)_{ij}=\vec{u}^\dagger_i\vec{u}_j$. To solve Eq. (\ref{ge}), the first step is the diagonalization of $N_m$ and one has
\begin{eqnarray}\label{dn}
\sum_{j=1}^m(N_m)_{ij}R^k_j=n_kR^k_i,
\end{eqnarray}
 where $n_k>0$ and $R^k$ with $k=1,2,...m$ are eigenvalues and the corresponding eigenvectors, respectively. Then one can establish a new set of orthonormal vectors  $\vec{v}_1,\vec{v}_2,...\vec{v}_m$ in the space spanned by $\vec{u}_1,\vec{u}_2,...\vec{u}_m$
 \begin{eqnarray}
 \vec{v}_k=\sum_{i=1}^m \frac {R^k_i\vec{u}_i}{\sqrt{n_k}}.
 \end{eqnarray}
It is easy to prove that the eigenvalues, $\lambda_k$, in Eq.(\ref{ge}) are identical to those of the Hermitian matrix $C_m$ with the elements $(C_m)_{ij}=v^\dagger_i(A_n)v_j$. According to the Poincar\'{e} separation theorem, we still have inequalities (\ref{ineq2}).

In the practical shell model calculations, the Lanczos algorithm is adopted. The Lanczos matrix is enlarged with the iterations. With Cauchy's interlacing theorem, one can easily understand why all the calculated energies decrease monotonically until they converge to the full shell model energies.

Let us denote the full shell model energies for a given nuclear Hamiltonian by $e_1\leq e_2\cdots$. Suppose that we are only interested in the lowest $m$ nuclear states. In all approximated shell model methods, efforts have been made to try to find a proper configuration subspace, so that the calculated lowest $m$ states are as close to the exact shell model ones as possible. Denoting the lowest $m$ approximated energies in a configuration subspace by $E_1\leq E_2\leq\cdots\leq E_m$, then it is always true that
\begin{eqnarray}\label{cauchy}
E_j\geq e_j \quad(1\leq j \leq m),
\end{eqnarray}
according to Cauchy's interlacing theorem or Poincar\'{e} separation theorem.

Therefore, one can define the non-negative energy differences $\delta E_j=E_j-e_j$ and the total energy difference
\begin{eqnarray}\label{DE}
\Delta E=\sum_{j=1}^m \delta E_j=\sum_{j=1}^m E_j-\sum_{j=1}^m e_j.
\end{eqnarray}
It is obvious that if $\Delta E=0$, then $\delta E_j=0$ for all calculated states and we have the exact shell model results. However, in the approximated methods, $\Delta E>0$, and one may expect that it should be as small as possible by adjusting the configuration subspace, in which the Hamiltonian is diagonalized. If $\Delta E$ reaches a minimum, then all the corresponding $m$ lowest energies are determined, and their approximation can be tested by comparing them with the shell model results.

Since $e_j$ energies in Eq.(\ref{DE}) are fixed for a given Hamiltonian, instead of minimizing the $\Delta E$, one can equivalently minimize the sum of the $m$ lowest $E_j$ energies, $S_m\equiv\sum_{j=1}^m E_j$. This makes the calculations more practical in the case of large model space, in which $e_j$ energies are impossible to be obtained.

Generally, it is unnecessary to restrict the form of the approximated nuclear wavefunctions. One may take the HF SDs, or HFB vacua with particle number projection, or their spin projected states, to form the configuration subspace. As a testing example of the present algorithm, here we take $n$ different HF SDs differed by $k$, $|\Phi^k\rangle$, in the model space that the Hamiltonian is defined, and project them onto good spin. These projected states then form a configuration subspace. The form of the nuclear states with good spin $J$ in that subspace can be written as
\begin{eqnarray}\label{wf}
|\Psi_{JM}\rangle=\sum_{k=1}^n\sum_{K=-J}^Jf_{Kk}P^J_{MK}|\Phi^k\rangle,
\end{eqnarray}
where, $P^J_{MK}$ is the spin (or angular momentum) projection operator, and $n$ is the number of adopted $|\Phi^k\rangle$ Slater determinants. $|\Phi^k\rangle$ can be established through the Thouless formula \cite{tuya17,Ring80},
\begin{eqnarray}\label{thouless}
|\Phi^k\rangle=\mathcal{N}_ke^{\frac12\sum_{\mu\nu}d^k_{\mu\nu}\beta^{k\dagger}_{0,\mu}\beta^{k\dagger}_{0,\nu}}|\Phi^k_0\rangle,
\end{eqnarray}
where $\mathcal{N}_k$ is the normalization parameter, and $\beta^{k\dagger}_{0,\mu}$ is the quasiparticle creation operator for the $|\Phi^k_0\rangle$ vacuum. The initial $|\Phi^k_0\rangle$ SDs can be obtained in the same way as in Ref. \cite{tuya17}, and they are assumed to be different from each other. Once $|\Phi^k_0\rangle$ SDs are randomly chosen, they no longer change. Thus $|\Phi^k\rangle$ can be conveniently varied only by changing the $d^k_{\mu\nu}$ parameters. Like Ref. \cite{tuya17}, we also assume the matrix elements, $d^k_{\mu\nu}$, to be complex numbers, i.e.,
\begin{eqnarray}
d^k_{\mu\nu}=x^k_{\mu\nu}+iy^k_{\mu\nu},
\end{eqnarray}
where, $x^k_{\mu\nu}$ and $y^k_{\mu\nu}$ are real numbers. For simplicity, we use $\vec{d}$ to denote the vector of all the independent $x^k_{\mu\nu}$ and $y^k_{\mu\nu}$ parameters.
With a given $\vec{d}$, all the $|\Phi^k\rangle$ SDs are determined and one can establish the Hill-Wheeler equation
\begin{eqnarray}\label{hw1}
\sum_{Kk}(H_{K'k', Kk}-EN_{K'k', Kk}) f_{Kk}=0
\end{eqnarray}
and the normalization condition
\begin{eqnarray}\label{hw2}
\sum_{K'k', Kk}f^*_{K'k'}f_{Kk}N_{K'k', Kk}=1,
\end{eqnarray}
where,
\begin{eqnarray}\label{hkk}
H_{K'k', Kk}&=&{\langle\Phi^{k'}|\hat{H}P^J_{K' K}|\Phi^k\rangle},\\\label{nkk}
N_{K'k', Kk}&=&{\langle\Phi^{k'}|P^J_{K' K}|\Phi^k\rangle}.
\end{eqnarray}
Solving Eq.(\ref{hw1}), one can get a set of energies $E_1\leq E_2\leq \cdots$ and the corresponding states $|\Psi^1_{JM}\rangle,|\Psi^2_{JM}\rangle,\cdots$.
Clearly, those energies are the functions of $\vec{d}$. Thus the sum of the lowest $m$ energies, $S_m$, is also determined by $\vec{d}$, and we can write $S_m=S_m(\vec{d})$.

Therefore, one can minimize $S_m$ by changing $\vec{d}$. For simplicity, we still call such energy minimization as variation after projection (VAP). To make the minimization efficient, the gradient of $S_m(\vec{d})$ should be calculated,
\begin{eqnarray}\label{grad}
\nabla S_m(\vec{d})=\sum_{j=1}^m\nabla E_j(\vec{d}),
\end{eqnarray}
where, the components of $\nabla E_j$ are actually $\frac{\partial E_j}{\partial x^k_{\mu\nu}}$ and $\frac{\partial E_j}{\partial y^k_{\mu\nu}}$, whose expressions can be easily deduced based on the formulas in Refs. \cite{tuya17,hu14}.

Once the quantities of $S_m(\vec{d})$ and its gradient $\nabla S_m(\vec{d})$ are available, one might perform the minimization of $S_m(\vec{d})$ using the quasi-Newton method. This may work if the Hessian at the minimum is positive definite. Unfortunately, we have learned that the Hessian of the projected energy at the the minimum is actually semi-positive definite \cite{tuya17}. This might be the reason that the traditional quasi-Newton method does not work well in our VAP calculations. To make the VAP calculation converges more reliably, we calculated the exact Hessian matrix, denoted by $\mathbf{H}[S_m(\vec{d})]$. Similar to Eq.(\ref{grad}), we also have
\begin{eqnarray}
\mathbf{H}[S_m(\vec{d})]=\sum_{j=1}^m \mathbf{H}[E_j(\vec{d})],
\end{eqnarray}
where the matrix elements of $\mathbf{H}[E_j(\vec{d})]$ are $\frac{\partial^2 E_j}{\partial x^k_{\mu\nu}\partial x^{k'}_{\mu'\nu'}}$, $\frac{\partial^2 E_j}{\partial x^k_{\mu\nu}\partial y^{k'}_{\mu'\nu'}}$ and $\frac{\partial^2 E_j}{\partial y^k_{\mu\nu}\partial y^{k'}_{\mu'\nu'}}$. The expressions for such second derivatives can also be derived without much difficulty based on the deductions in Refs. \cite{tuya17,hu14}.

To check the correctness of the above calculated quantities, one can use the following equations by definition,
\begin{eqnarray}
&&\lim_{\delta \vec{d}\rightarrow0} \frac{S_m(\vec{d}+\delta \vec{d})-S_m(\vec{d})}{|\delta \vec{d}|}=\nabla S_m(\vec{d})\cdot\vec{e},\label{d1}\\
&&\lim_{\delta \vec{d}\rightarrow0} \frac{\nabla S_m(\vec{d}+\delta \vec{d})-\nabla S_m(\vec{d})}{|\delta \vec{d}|}\cdot\vec{e}\nonumber\\
&=&\vec{e}^t\cdot\mathbf{H}[S_m(\vec{d})]\cdot\vec{e},\label{d2}
\end{eqnarray}
where, $\vec{e}=\frac{\delta\vec{d}}{|\delta \vec{d}|}$, which shows the direction of $\delta \vec{d}$. In the present numerical calculations, the calculated $S_m$, its gradient and Hessian matrix indeed fulfill the above equations at arbitrary point $\vec{d}$ and $\vec{e}$. With a small $|\delta \vec{d}|=10^{-4}$, the calculated differences between both sides of Eqs. (\ref{d1}) and (\ref{d2}) are usually less than $10^{-6}$.

\begin{figure}
  \includegraphics[width=3.4in]{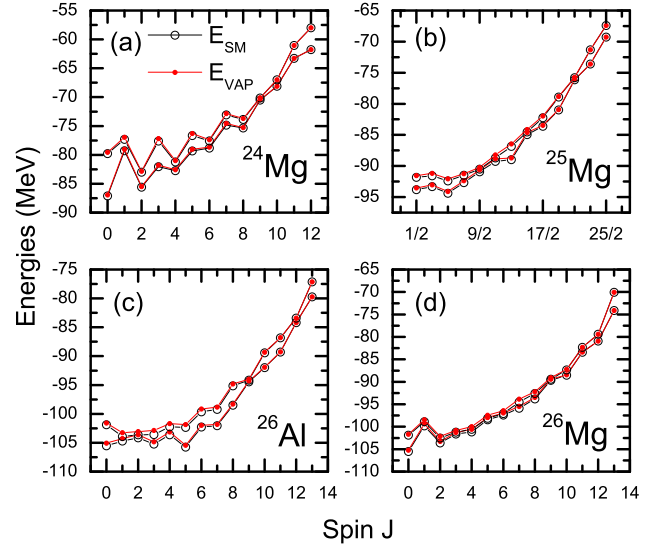}\\
  \caption{ (Color online)  Calculated two lowest VAP energies with two Slater determinants, $E_{VAP}$, and the corresponding shell model energies, $E_{SM}$, as functions of spin $J$  for (a) $^{24}$Mg , (b) $^{25}$Mg, (c) $^{26}$Al, and (d) $^{26}$Mg. The USDB interaction is adopted.}\label{re}
\end{figure}

Once the above quantities are available, we start from $\vec{d}=0$ and take the exact trust region method \cite{numopt} to search for the minimum of $S_m(\vec{d})$.  At the minimum, the gradient $\nabla S_m(\vec{d})$ should be zero. Here, the VAP iteration terminates if $|\nabla S_m(\vec{d})|\leq 0.01$keV. This is a very strict condition so that the obtained minimum is precise.

\begin{figure}
  \includegraphics[width=3.4in]{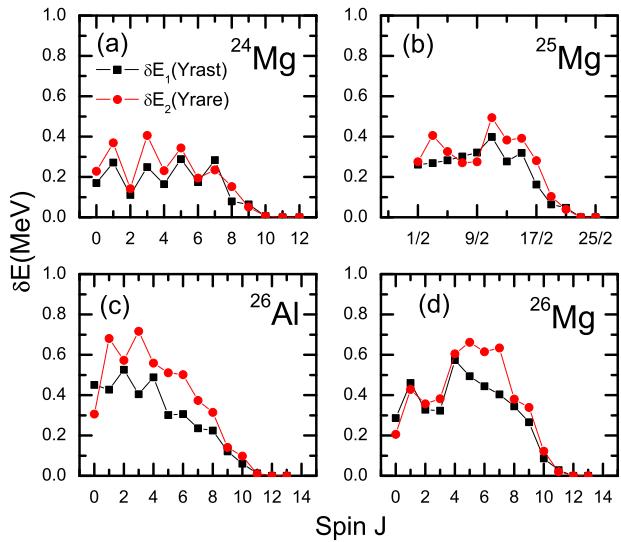}\\
  \caption{ (Color online)  Energy differences between the shell model energies, $E_{SM}$, and the present VAP energies, $E_{VAP}$, in Fig.\ref{re}.}\label{ediff}
\end{figure}

To test the validity of the present algorithm, we performed the VAP calculations in the $sd$ shell and take the USDB interaction \cite{usdb}. The simplest calculation for the non-yrast states is the one with $n=2$ and $m=2$, i.e., at a given spin, we take two SDs to calculate the lowest two nuclear states, simultaneously. The calculated nuclei are $^{24}$Mg, $^{25}$Mg, $^{26}$Mg, and $^{26}$Al, whose calculated energies are shown in Fig.\ref{re}. It looks that all the calculated VAP energies are very close to the shell model ones calculated with the NuSHellX\cite{nushellx}. To show the differences between the VAP and the shell model more clearly, we calculated the $\delta E_j$ $(j=1,2)$ from Fig.\ref{re}, and show them in Fig.\ref{ediff}. One can see that all the $\delta E_j$ are indeed nonnegative, as predicted by Cauchy's interlacing theorem. It is interesting to see that the energy differences of the yrare states, $\delta E_2$, are roughly close to the yrast ones, $\delta E_1$ after the minimization of $S_{m=2}$.

\begin{figure}
  \includegraphics[width=3.4in]{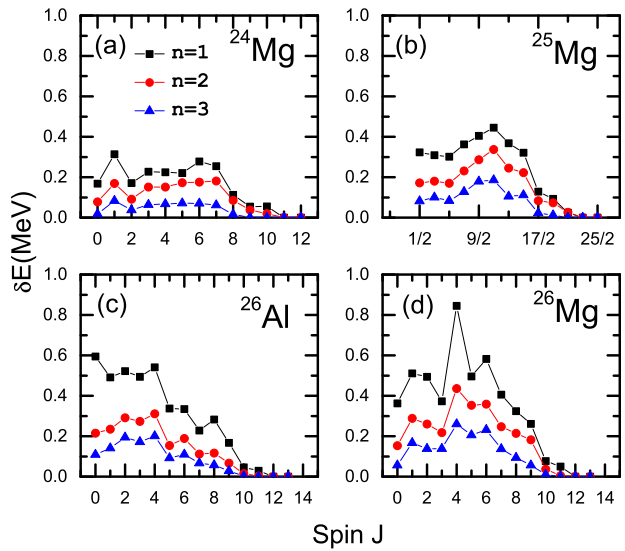}\\
  \caption{ (Color online)  Energy differences $\delta E_1$ with $m=1$ and $n=1,2$ and 3, as functions of the spin $J$.}\label{e3diff}
\end{figure}

 The $\delta E_i$ values are expected to be as small as possible so that one can obtain satisfactory
  shell model approximations. This can be realized by adding more SDs to the VAP wave functions.
   As a testing example, we only calculated the yrast states (i.e., $m=1$) with the cases $n=1,2$ and 3.
 The calculated $\delta E_1$ energy differences are shown in Fig.\ref{e3diff}. The case with $n=1$ is exactly the Fig.2 in Ref.\cite{tuya17}. For the $n=2$ case, all the $\delta E_1$ values (except for the ones at the two highest spins) are smaller than the corresponding ones with $n=1$. Notice that for the two highest spin states, we have $\delta E_1=0$ in all cases due to the unphysical tiny configuration space. With more SDs added in, the VAP results become more closer to the shell model ones.
 One can see that, again, the $\delta E_1$ with $n=3$ becomes even smaller.

\begin{figure}
  \includegraphics[width=3.4in]{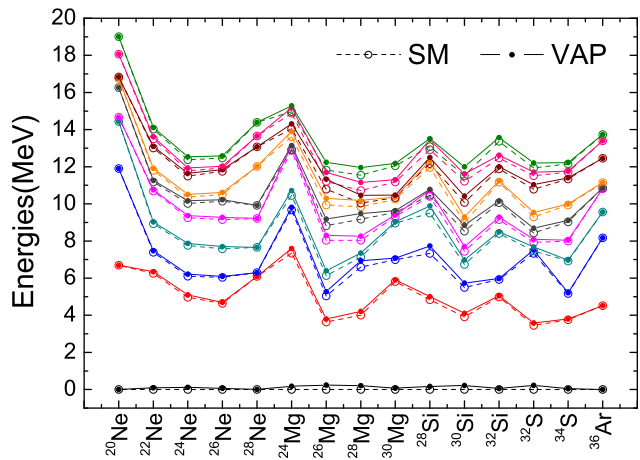}\\
  \caption{ (Color online)  Calculated 10 lowest $J^\pi=0^+$ VAP energies with 10 SDs for the even-even $sd$ shell nuclei
  and the corresponding shell model energies. The shell model ground states  are set to zero.}\label{10}
\end{figure}

However, in the practical application, it may not be enough that only two lowest states are calculated. Here, to show the validity of the present algorithm, we calculated the lowest 10 states ($m=10$) with 10 SDs ($n=10$) for the $J=0$ states in even-even $sd$ shell nuclei. The results are shown in Fig.\ref{10}. Like Fig.\ref{re}, the VAP energies are also very close to the corresponding SM ones. Again, all the $\delta E_i$ are non-negative. Especially for the $^{20}$Ne, $^{28}$Ne and $^{36}$Ar nuclei, we have $\delta E_i=0$ for all the calculated states. This means we got the exact shell model results for these 3 nuclei and confirms the correctness of the present calculations. Most of  $\delta E_i$ values are in the range from 0 to 300keV. Some of them are a little larger but the largest one is still within 600keV. Of course, such approximation can be further improved by increasing the number, $n$, of the included SDs.

It may be interesting to compare the present algorithm with the one taken by the VAMPIR. Let us first calculate a simplest case of $m=1$ and $n=2$ for the ground $0^+$ state in $^{24}$Mg. The USDB interaction is still adopted and the exact ground state energy in $^{24}$Mg is $-87.104$MeV. We vary these two SDs simultaneously in the present VAP and obtained the lowest minimum $-87.039$MeV. Now, let us follow the method of FED VAMPIR \cite{Schmid04}.  Varying the first SD,  one can get the energy  minimum of $-86.936$MeV. Then a second SD is added to improve the approximation. Fixing the obtained first SD and varying the second one, this makes the ground state energy lower to $-87.008$MeV which still lies above our $-87.039$MeV. This clearly shows the SDs obtained by the FED VAMPIR algorithm are not fully optimized. As can be seen that if one fixes the second SD and comes back to vary the first SD again, then the ground state energy drops further from $-87.008$MeV to $-87.015$MeV.

Another comparison is the case of $m=2$ and $n=2$ for the ground $0^+$ state and the first excited $0^+$ state in $^{24}$Mg. Our algorithm gives the lowest $S_2$ to be $-166.474$MeV which corresponds to the lowest two $0^+$ energies $E_1=-86.944$MeV and $E_2=-79.530$MeV. While the exact SM results are $e_1=-87.104$MeV and $e_2=-79.766$MeV. If we follow the algorithm in the EXCITED VAMPIR \cite{Schmid04}, then the ground state energy obtained by varying the first SD is $E_1=-86.936$MeV as mentioned in the last paragraph. For the minimization of $E_2$, the first SD will be fixed and the second SD will be varied under the Gram-Schmidt orthogonalization. This is essentially equivalent to the minimization of $S_2=E_1+E_2$ because $E_1$ is fixed. After the minimization of $E_2$, the VAMPIR performs the final dialgonalization in the space spanned by the two optimized projected SD states and the final lowest two energies can be obtained. In the present case of $J^\pi=0^+$, such dialgonalization does not change $S_2$. Therefore, it is clear that the whole procedure of the EXCITED VAMPIR is exactly equivalent to the minimization of $S_2$ in our VAP but with the first SD fixed. In this way, the obtained $S_2=166.341$MeV, corresponding to $E_1=-86.940$MeV and $E_2=-79.400$MeV. Both energies are above the present VAP results, especially $E_2$, which is $130$keV above our $-79.530$MeV. This again shows that the projected states in the VAMPIR method are not fully optimized.

\begin{figure}
  \includegraphics[width=3.4in]{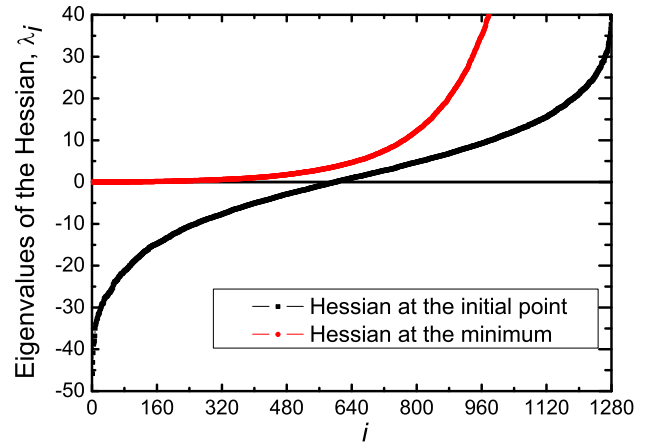}\\
  \caption{ (Color online)  Eigenvalues of the Hessian matrix corresponding to the VAP calculation of $^{24}$Mg in Fig.\ref{10}. The black rectangles show the ones at $\vec{d}=0$. The red dots show the ones after $S_{m=10}$ converges to a minimum. The number of total VAP parameters for $^{24}$Mg is 1280. Each SD takes 128 VAP paramters, see Ref.\cite{tuya17}.}\label{hs}
\end{figure}

Let us study the Hessian matrix by calculating its eigenvalues. In Fig.\ref{hs}, we show the eigenvalues of the Hessian matrix corresponding to the VAP calculations of $^{24}$Mg in Fig.\ref{10}. Clearly, the Hessian matrix is indefinite at the starting point. But after convergence, it becomes semipositive definite. This is very similar to the case of $m=1$ and $n=1$ as discussed in Ref.\cite{tuya17}.

Due to the simplicity of the present algorithm, all the above VAP calculations are performed with a common code, except for taking different $m,n (n\geq m)$ and the spin $J$ as input parameters. We hope the present VAP will be an alternative way of extending the shell model calculations in larger model spaces. Calculations for heavier nuclei will be performed after our VAP code is parallelized.

 Equation (\ref{cauchy}) is essentially a mathematical conclusion. The present method of the $S_m$ minimization may be universal and applicable to other quantum many-body systems; for instance, the excited electronic states in a chemical system \cite{carlos13}. One may also change the form of the approximated wavefunction in a convenient way. For example, if one simply takes the superposition of the deformed SDs without spin projection, then the minimization of $S_m$ should be a natural extension of the Hartree-Fock method for the excited states. Such calculation can be implemented by simply removing the spin projection from the present VAP code.

\textbf{Acknowledgements} The work is supported by the
National Natural Science Foundation of China under Contract Nos.
11575290, 11475072 and 11490560.






\end{document}